\documentstyle[prd,aps,12pt,epsf,psfig]{revtex}
\begin{document}

\baselineskip=7mm
\def\ap#1#2#3{           {\it Ann. Phys. (NY) }{\bf #1} (#2) #3}
\def\arnps#1#2#3{        {\it Ann. Rev. Nucl. Part. Sci. }{\bf #1} (#2) #3}
\def\cnpp#1#2#3{        {\it Comm. Nucl. Part. Phys. }{\bf #1} (#2) #3}
\def\apj#1#2#3{          {\it Astrophys. J. }{\bf #1} (#2) #3}
\def\asr#1#2#3{          {\it Astrophys. Space Rev. }{\bf #1} (#2) #3}
\def\ass#1#2#3{          {\it Astrophys. Space Sci. }{\bf #1} (#2) #3}

\def\apjl#1#2#3{         {\it Astrophys. J. Lett. }{\bf #1} (#2) #3}
\def\ass#1#2#3{          {\it Astrophys. Space Sci. }{\bf #1} (#2) #3}
\def\jel#1#2#3{         {\it Journal Europhys. Lett. }{\bf #1} (#2) #3}

\def\ib#1#2#3{           {\it ibid. }{\bf #1} (#2) #3}
\def\nat#1#2#3{          {\it Nature }{\bf #1} (#2) #3}
\def\nps#1#2#3{          {\it Nucl. Phys. B (Proc. Suppl.) } {\bf #1} (#2) #3}
\def\np#1#2#3{           {\it Nucl. Phys. }{\bf #1} (#2) #3}

\def\pl#1#2#3{           {\it Phys. Lett. }{\bf #1} (#2) #3}
\def\pr#1#2#3{           {\it Phys. Rev. }{\bf #1} (#2) #3}
\def\prep#1#2#3{         {\it Phys. Rep. }{\bf #1} (#2) #3}
\def\prl#1#2#3{          {\it Phys. Rev. Lett. }{\bf #1} (#2) #3}
\def\pw#1#2#3{          {\it Particle World }{\bf #1} (#2) #3}
\def\ptp#1#2#3{          {\it Prog. Theor. Phys. }{\bf #1} (#2) #3}
\def\jppnp#1#2#3{         {\it J. Prog. Part. Nucl. Phys. }{\bf #1} (#2) #3}

\def\rpp#1#2#3{         {\it Rep. on Prog. in Phys. }{\bf #1} (#2) #3}
\def\ptps#1#2#3{         {\it Prog. Theor. Phys. Suppl. }{\bf #1} (#2) #3}
\def\rmp#1#2#3{          {\it Rev. Mod. Phys. }{\bf #1} (#2) #3}
\def\zp#1#2#3{           {\it Zeit. fur Physik }{\bf #1} (#2) #3}
\def\fp#1#2#3{           {\it Fortschr. Phys. }{\bf #1} (#2) #3}
\def\Zp#1#2#3{           {\it Z. Physik }{\bf #1} (#2) #3}
\def\Sci#1#2#3{          {\it Science }{\bf #1} (#2) #3}

\def\n.c.#1#2#3{         {\it Nuovo Cim. }{\bf #1} (#2) #3}
\def\r.n.c.#1#2#3{       {\it Riv. del Nuovo Cim. }{\bf #1} (#2) #3}
\def\sjnp#1#2#3{         {\it Sov. J. Nucl. Phys. }{\bf #1} (#2) #3}
\def\yf#1#2#3{           {\it Yad. Fiz. }{\bf #1} (#2) #3}
\def\zetf#1#2#3{         {\it Z. Eksp. Teor. Fiz. }{\bf #1} (#2) #3}
\def\zetfpr#1#2#3{         {\it Z. Eksp. Teor. Fiz. Pisma. Red. }{\bf #1} (19#2) #3}
\def\jetp#1#2#3{         {\it JETP }{\bf #1} (19#2) #3}
\def\mpl#1#2#3{          {\it Mod. Phys. Lett. }{\bf #1} (#2) #3}
\def\ufn#1#2#3{          {\it Usp. Fiz. Naut. }{\bf #1} (#2) #3}
\def\sp#1#2#3{           {\it Sov. Phys.-Usp.}{\bf #1} (#2) #3}
\def\ppnp#1#2#3{           {\it Prog. Part. Nucl. Phys. }{\bf #1} (#2) #3}
\def\cnpp#1#2#3{           {\it Comm. Nucl. Part. Phys. }{\bf #1} (#2) #3}
\def\ijmp#1#2#3{           {\it Int. J. Mod. Phys. }{\bf #1} (#2) #3}
\def\ic#1#2#3{           {\it Investigaci\'on y Ciencia }{\bf #1} (#2) #3}
\def\tp{these proceedings}
\def\pc{private communication}
\def\ip{in preparation}
\newcommand{\TeV}{\,{\rm TeV}}
\newcommand{\GeV}{\,{\rm GeV}}
\newcommand{\MeV}{\,{\rm MeV}}
\newcommand{\keV}{\,{\rm keV}}
\newcommand{\eV}{\,{\rm eV}}
\newcommand{\Tr}{{\rm Tr}\!}
\renewcommand{\arraystretch}{1.2}
\newcommand{\be}{\begin{equation}}
\newcommand{\ee}{\end{equation}}
\newcommand{\bea}{\begin{eqnarray}}
\newcommand{\eea}{\end{eqnarray}}
\newcommand{\ba}{\begin{array}}
\newcommand{\ea}{\end{array}}
\newcommand{\bmat}{\left(\ba}
\newcommand{\emat}{\ea\right)}
\newcommand{\refs}[1]{(\ref{#1})}
\newcommand{\ler}{\stackrel{\scriptstyle <}{\scriptstyle\sim}}
\newcommand{\ger}{\stackrel{\scriptstyle >}{\scriptstyle\sim}}
\newcommand{\lag}{\langle}
\newcommand{\rag}{\rangle}
\newcommand{\ns}{\normalsize}
\newcommand{\cm}{{\cal M}}
\newcommand{\gr}{m_{3/2}}
\newcommand{\p}{\partial}
\renewcommand{\le}{\left(}
\newcommand{\ri}{\right)}
\renewcommand{\o}{\overline}

\relax
\def\321{$SU(3)\times SU(2)\times U(1)$}
\def\21{$SU(2)\times U(1)$}
\def\dbd{$0\nu\beta\beta~$}
\def\ord{{\cal O}}
\def\tl{{\tilde{l}}}
\def\tL{{\tilde{L}}}
\def\bd{{\overline{d}}}
\def\tL{{\tilde{L}}}
\def\a{\alpha}
\def\b{\beta}
\def\g{\gamma}
\def\c{\chi}
\def\d{\delta}
\def\D{\Delta}
\def\db{{\overline{\delta}}}
\def\Db{{\overline{\Delta}}}
\def\e{\epsilon}
\def\l{\lambda}
\def\n{\nu}
\def\m{\mu}
\def\nt{{\tilde{\nu}}}
\def\p{\phi}
\def\P{\Phi}
\def\solm{\Delta_{\odot}}
\def\sola{\theta_{\odot}}
\def\mee{m_{ee}}
\def\atm{\Delta_{\makebox{\tiny{\bf atm}}}}
\def\k{\kappa}
\def\x{\xi}
\def\r{\rho}
\def\s{\sigma}
\def\t{\tau}
\def\th{\theta}
\def\om{\omega}
\def\ne{\nu_e}
\def\nm{\nu_{\mu}}
\def\snui{\tilde{\nu_i}}
\def\ehat{\hat{e}}
\def\la{{\makebox{\tiny{\bf loop}}}}
\def\ta{\tilde{a}}
\def\tb{\tilde{b}}
\def\mb{m_{1b}}
\def\mt{m_{1 \tau}}
\def\rl{{\rho}_l}
\def\meg{\m \rightarrow e \g}

\renewcommand{\Huge}{\Large}
\renewcommand{\LARGE}{\Large}
\renewcommand{\Large}{\large}
\title{\large \bf Radiatively Generated $\nu_e$ oscillations:\\General
Analysis,
Textures and Models}
\author{ Anjan S. Joshipura and Saurabh D. Rindani\\[.5cm]
{\ns\it  Theoretical Physics Group, Physical Research Laboratory,}\\
{\ns\it Navrangpura, Ahmedabad 380 009, India.}}
\date{}
\maketitle
\vskip .5cm
\begin{center}
{\bf Abstract}
\end{center}
We study the consequences of assuming that the mass scale $\solm$
corresponding to the solar neutrino oscillations and mixing angle
$U_{e3}$ corresponding to the electron neutrino oscillation at
CHOOZ are radiatively generated through the standard electroweak
gauge interactions. All the leptonic mass matrices having zero
$\solm$ and $U_{e3}$ at a high scale lead to a unique low energy
value for the $\solm$  which is determined by the (known) size
of the radiative corrections, solar and the atmospheric mixing
angle and the Majorana mass of the neutrino observed in 
neutrinoless double beta decay. This prediction leads to the
following consequences: ($i$) The MSSM radiative corrections
generate only the dark side of the solar neutrino solutions.
($ii$) The inverted mass hierarchy ($m,-m,0$)  at the high scale
fails in generating the LMA solution but it can lead to the LOW or
vacuum solutions. ($iii$) The $\solm$ generated in models with
maximal solar mixing at a high scale is zero to the lowest order
in the radiative parameter. It tends to get suppressed as a result
of this and lies in the vacuum region. We discuss specific textures
which can lead to the LMA solution in the present framework and
provide a gauge theoretical realization of this in the context of
the seesaw model.
\newpage

\section{Introduction} The large amount of information gathered from
solar and atmospheric neutrino experiments provides a detailed
probe of the neutrino mass spectrum \cite{sf}. The quark-lepton
symmetry implemented in the seesaw model for neutrino masses
prefers hierarchical neutrino masses \cite{revs}. This pattern can
easily explain the mass scales involved in neutrino oscillations,
but obtaining two large mixing angles required phenomenologically
needs considerable changes in the simple seesaw schemes. Still,
many schemes can accommodate large mixing in the seesaw picture
\cite{sm,barr}.

The alternative to a hierarchical neutrino spectrum is the
presence of a pseudo-Dirac state corresponding to two nearly
degenerate neutrinos with splitting much smaller than their
overall mass scale. The third neutrino may be much lighter than
this mass (leading to inverted hierarchy) or can be close to it
giving an almost degenerate spectrum \cite{deg}. Unlike in the
hierarchical spectrum, large mixing angles can be quite natural in
such schemes and various models \cite{barb,emt,bm,bim} predicting
these mass patterns also predict large or maximal mixing angles.

There are two physically different zeroth order realizations of a
pseudo-Dirac neutrino in the case of two generations, say, $e$ and
$\m$. These are given by the following mass matrices:
\be \label{dirac} \left( \ba{cc} 0&m\\m&0\\\ea \right) \ee
and
\be \label{pd2} \left( \ba{cc} a&b\\b&-a\\
\ea \right) \ee

Both these matrices give equal and opposite eigenvalues at the
tree level. The one in eq.~(\ref{dirac}) is invariant under the
$L_e-L_\m$ symmetry. The neutrinos remain as a Dirac particle as
long as this symmetry remains exact but its breaking would lead to
splitting  among neutrinos. The $L_e-L_\m$ symmetry cannot be
broken by the standard electroweak interactions and one needs to
invoke additional interactions to obtain a Dirac neutrino from
eq.~(\ref{dirac}). The matrix of eq.~(\ref{pd2}) is quite
different. The mass eigenstates emerging from eq.~(\ref{pd2}) also
form a Dirac neutrino pair due to equal and opposite eigenvalues
but the charged current interactions defined in this mass basis of
the degenerate pairs violate lepton number \cite{wolf}. As a
consequence, the standard electroweak interactions radiatively
generate neutrino splitting unlike in the case of
eq.~(\ref{dirac}). Thus the matrix in eq.~(\ref{pd2})
intrinsically defines a pseudo-Dirac pair. Both the matrices also
differ phenomenologically since the one in eq.~(\ref{dirac}) leads
to practically maximal mixing while the mixing angle is arbitrary
in the case of eq.~(\ref{pd2}).

We can also divide pseudo-Dirac neutrinos into two different
categories on phenomenological grounds, one in which the
effective neutrino mass probed through neutrinoless double beta
decay (\dbd) is comparable to the mass of the neutrino pair, and
the other in which it is much smaller. These two cases are exemplified
by eq.~(\ref{pd2}) and eq.~(\ref{dirac}) respectively.

We need  $3\times 3$ generalizations of eq.~(\ref{pd2}) for
phenomenological purposes. It is possible to write such
generalizations \cite{pdp,pdp1} and to go even one step further
and classify all the allowed forms of neutrino mass matrices which
lead to a pseudo-Dirac neutrino after inclusion of the the
standard weak radiative corrections. Conditions for this to happen
were studied in \cite{pdp}. General solutions of these conditions
in CP conserving theories were worked out \cite{pdp2} under an
assumption that the element $U_{e3}$ of the neutrino mixing matrix
$U$ is zero at a high scale. Both the solar scale and the $U_{e3}$
get generated at a low scale $\m$ after radiative corrections are
included. But these quantities get related to low energy
observables \cite{pdp2}. Specifically, one gets the following
unique prediction in this case:
\be\label{pred} \solm \cos 2\sola=4 \delta_{\tau} \sin^2\theta_A
|m_{ee}|^2 +\ord(\delta^2_{\tau})~. \ee

Here $\solm$ is the mass-squared difference responsible for the solar neutrino
oscillation, and the angles $\sola$ and $\theta_A$ respectively denote the solar
and atmospheric mixing angles at a low scale. $m_{ee}$ is the
effective neutrino mass probed in the \dbd decay and
$\delta_\tau$ specifies the size of the radiative correction induced
by the Yukawa coupling of the $\tau$:
\be \label{delta1}
\delta_\tau\approx c\left({m_\tau\over 4 \pi v }\right)^2 \ln{M_X\over M_Z}~.
\ee
$c=\frac{3}{2},-\frac{1}{\cos^2\b}$ in the case of the standard
model (SM), and  the minimal supersymmetric standard model (MSSM),
respectively \cite{rg,radrev,rad}.

The prediction in eq.~(\ref{pred}) is independent of the detailed
form of the neutrino and charged-lepton mass matrices at a high
scale and holds also in the presence of CP violation \cite{cpv}.
The only assumption is that the leptonic mass matrices yield a
degenerate neutrino pair and vanishing $U_{e3}$ at a high scale.
This remarkably general relation  shows a strong correlation
between the solar scale and $m_{ee}$. In particular, the LMA
solution requires \cite{pdp2} $m_{ee}$ to be close to the present
experimental limit \cite{limit}.

In this paper, we study consequences of the general picture
discussed above and  discuss some specific textures and their
gauge theoretical realizations. Our basic assumptions are (a) CP
conservation (b) vanishing $U_{e3}$ at a high scale (c) neutrino
masses with high-scale eigenvalues $m,-m,m'$ (d) the presence of
only standard weak radiative corrections either in SM or in MSSM.
Three general consequences follow from these assumptions: ($i$)
The inverted mass hierarchy among neutrinos is inconsistent with
the large mixing angle solution (LMA) to the solar neutrino
problem but the LOW solution is allowed by it. ($ii$) The
assumption of maximal solar mixing angle at a high scale also
fails in reproducing the solar scale required for either the LMA
or the LOW solution in SM but it can account for the $\solm$
corresponding to the VAC solution. ($iii$) Radiative corrections
in MSSM cannot reproduce any solution of the solar neutrino
problem which require $\tan^2\sola\leq 1$.

The next section is devoted to a discussion of our basic scheme and
the above mentioned general results. Then we discuss some specific
high scale textures for the neutrino and charged-lepton mass
matrices which are phenomenologically viable. We describe in
Section 4 a realization of one of these textures in the context
of the standard seesaw model augmented with a horizontal
$SU(2)\times U(1)$ symmetry.

\section{Consequences of a pseudo-Dirac structure}
We first summarize the basic framework and main results of ref
\cite{pdp,pdp2} and then discuss their consequences.

Consider a CP conserving theory specified by a general $3\times 3$
real symmetric neutrino mass matrix $M_{\nu 0}$. This matrix can
always be specified in the flavor basis corresponding to a
diagonal charged-lepton mass matrix. We adopt the following
parameterization for $M_{\nu 0}$ in this basis.
\be \label{para}M_{\nu 0}= \bmat{ccc} s_1&t&u\\
t&s_2&v\\u&v&s_3\\ \emat \ee
Assume that the above $M_{\nu 0}$ describes physics at a high
scale $M_X$. $M_{\nu 0}$ is required to yield vanishing solar
scale and CHOOZ angle at $M_X$. We require that $M_{\nu 0}$ has
eigenvalues $(m,-m,m')$. This happens if \cite{pdp}
\be \label{cond1}
 tr(M_{\nu 0})\sum_i \Delta_i=det M_{\nu 0} ~,\ee where $\Delta_i$
represents the determinant of the $2\times 2$ block of $M_{\nu 0}$
obtained by blocking the $i^{\rm th}$ row and column. When the above
condition is satisfied, eigenvalues of $M_{\nu 0}$ are given in
terms of invariants of $M_{\nu 0}$
\be \label{ev}
 m\equiv \sqrt{-\sum_i{\Delta_i}}~~~~~~;~~~~~~m'=T\equiv tr(M_{\nu
0}) \ee
A general solution of eq.~(\ref{cond1}) can be obtained if one
further assumes that $U_{e3}$ is zero at $M_X$. This solution
corresponds to
\bea \label{cond} v^2&=& (s_1+s_2)(s_1+s_3)~; \nonumber \\
t&=&-{u v \over s_1+s_2}~. \eea
The above equations ensure the degeneracy (up to a sign) of two
states and vanishing $U_{e3}$ at a high scale. Radiative
corrections to $M_{\nu 0 }$ can be incorporated in the standard
way \cite{radrev}. It was shown \cite{pdp,pdp2} that the
radiative corrections do not destabilize the basic structure.
Their main effect is to generate the solar scale $\solm$ and
$U_{e3}$. The $\solm$ obtained after radiative corrections is
related to $\mee$ by eq.~(\ref{pred}) for all neutrino mass
matrices whose elements (in the flavour basis) satisfy
eq.~(\ref{cond}). The other observables are given by
\bea\label{obs} \tan^2\theta_A&\approx&{u^2\over t^2}+\ord
(\delta_\tau),
\nonumber \\
\tan^2\sola&=&{m-\mee\over m+\mee}+\ord(\delta_{\tau}^2),
\nonumber \\
\Delta_A&\approx & |m^2-T^2|~. \eea
$m$ is the common mass of the degenerate pair\footnote{ We have
neglected here an inconsequential  numerical factor \cite{radrev}
corresponding to the overall renormalization of the neutrino mass
matrix in eq.~(\ref{para}).} and is given from
eq.~(\ref{ev},\ref{cond}) by
\be \label{m} m=\sqrt{\mee^2+t^2+u^2}\ee

All the results presented here are independent of the specific
structures for the charged-lepton and neutrino mass matrices. We use
these results now to discuss several interesting and fairly
model-independent consequences.

\subsection{Radiative corrections and MSSM}

The available solar neutrino results are known to significantly
constrain \cite{solb} the solar scale $\solm$ and the mixing
$\sola$, particularly after \cite{sola} the recent neutral current
results from SNO \cite{sno}. Based on the global analysis of all
the solar data, the only solutions allowed at 3$\sigma$ level are
the large mixing angle (LMA) solution and the LOW or quasi vacuum
\cite{smnu00}. The allowed ranges of parameters in these cases at
3$\sigma$ are given approximately by by $\solm\approx 3\cdot
10^{-4}-2 \cdot 10^{-5} \eV^2$ and $\tan^2\sola\sim 0.2-0.9$ in
case of the LMA solution and $\solm\approx 3 \cdot 10^{-8}- 1
\cdot 10^{-7} \eV^2$ and $\tan^2\sola\sim 0.4-0.9$ in case of the
LOW solution. The small mixing angle solution is excluded at
3$\sigma$. The solar mixing angle is found to be less than $45^0$
($\tan^2\sola\leq 0.84$) in all the preferred solutions, a fact
which plays an important role in the following.

 The
predicted values of $\solm$ and $U_{e3}$  are different in SM and
MSSM due to different values of  $\delta_\tau$ in these two cases.
In  case of the SM, $\delta_\tau\sim 10^{-5}$ while it can become
larger for the MSSM due to the presence of $\tan\b$. More
importantly, the sign of $\delta_{\tau}$ is different in these two
cases. The negative values of $\delta_{\tau}$ for MSSM makes it
unsuitable for the description of the solar data. This important
result is a direct consequence of the general prediction in
eq.(\ref{pred}). It implies that the $\solm \cos 2 \sola \leq 0$
as long as $m_{ee}$ is not suppressed\footnote{It may be possible
for the MSSM radiative corrections to account for the solar scale
through $\ord(\delta_\tau^2)$ corrections if $m_{ee}$ is close to
zero. We shall present an explicit example of this latter on.},
typically if $m_{ee}^2\geq m^2 \delta_{\tau}$. But the preferred
solar neutrino solutions LMA and LOW require a positive $\solm
\cos 2 \sola $. Hence these solutions cannot be realized in the
case of MSSM. The best fit point for the vacuum solution falls in
this dark zone \cite{smnu00} and hence this solution can in
principle be realized. This would require $|m_{ee}|^2\sim (m \cos
2\sola)^2 \leq 10^{-5} \eV^2$ since $\delta_{\tau} \geq 10^{-5}$
in case of the MSSM.

\subsection{Inverted mass hierarchy}
The pattern of neutrino masses $m_{\nu_i}$ corresponding to the
relation
$$ m_{\nu_1}\sim m_{\nu_2}\gg m_{\nu_3}$$
is termed as the inverted mass hierarchy. If the solar scale is
generated radiatively then the inverted mass pattern at the low
scale would correspond to the masses $(m,-m,m'\ll m)$ at the high
scale. A simple realization of this pattern is provided by the
$L_{e}-L_\m-L_\t$ symmetry \cite{emt,bm} which leads to neutrino
masses $(m,-m,0)$. The mass $m$ is identified with the atmospheric
scale and  a small breaking of the $L_{e}-L_\m-L_\t$ symmetry can
generate the solar scale. Such breaking can come from the standard
radiative corrections \cite{bm,barb,pdp} or from additional
interactions \cite{emt}. We now show that if the standard
radiative breaking is responsible for the generation of the solar
scale as in \cite{barb,bm} then one cannot obtain the LMA solution
of the solar neutrino problem in case of the inverted hierarchy.
Our conclusion is not restricted only to the $L_{e}-L_\m-L_\t$
symmetric models but holds in case of all the models with the
inverted mass pattern $(m,-m,m'\ll m)$ and vanishing $U_{e3}$ at a
high scale. It follows from eq.~(\ref{ev}) that  this pattern is
obtained in all the models in which the neutrino mass matrix has
vanishing or very small trace ($T\ll m$). Models with
$L_{e}-L_\m-L_\t$ symmetry provide one example of this. In all these
models, the common mass $m$ of the degenerate pair needs to be
identified with the atmospheric neutrino mass scale. Eq. (\ref{m})
then implies
$$ \mee^2\leq m^2\approx\Delta_A$$
showing that in models with inverted mass hierarchy the \dbd decay
scale can at most be of the order of the atmospheric scale. This
then restricts the allowed values of the solar scale due to the
strong correlation eq.~(\ref{pred}) that exists between the
$\solm$ and $\mee$.  The parameter $\delta_\tau$ appearing in
eq.~(\ref{pred}) is given by eq.~(\ref{delta1}). Since MSSM is
unable to reproduce the solar mixing, there is no arbitrariness in
the choice of $\delta_\t$ whose numerical value is fixed here to
be $\sim 3\times 10^{-5} $. The $\cos 2 \sola$ is required to be
in the range $0.08-0.6 $ for the LMA solution at $3 \sigma$.
Taking the extreme and the most favourable values for the
parameters within the allowed range, we obtain from
eq.~(\ref{pred})
\be \label{invert} \solm \leq 5  \times 10^{-6} \eV^2 ~, \ee where
we have chosen $\mee^2=\Delta_A=(.08)^2 \eV^2$. This value of
$\solm$ still falls short of the value required for the LMA
solution $\sim (2.3-37)\cdot 10^{-5}\eV^2$.

It follows that the inverted hierarchy cannot be reconciled with
the radiatively generated LMA solution. Although we assumed
zero $U_{e3}$ at a high scale, the correction
to the radiatively generated $\solm$ coming from a non-zero
$U_{e3}$ is expected to be $\sim \delta_{\tau} \atm U_{e3}$ since
there is only one mass scale corresponding to $m^2\sim \atm$ in
the inverted hierarchy scheme. This correction is smaller compared
to the LMA scale for $U_{e3}\sim 0.1$ and $\delta_{\tau}\sim
10^{-5}$. Thus inclusion of a non-zero $U_{e3}$ is unlikely to
change the above conclusion at least in the case of the standard
model.

\subsection{Maximal solar mixing at high scale}

The required large solar mixing angle has led to a hypothesis of a
maximal solar mixing angle. This together with the large
atmospheric  mixing leads to bi-maximal mixing pattern which has
been studied quite extensively \cite{bim}. This possibility is
theoretically attractive \cite{nsg} since maximal mixing can be
obtained by imposing some symmetry. The present solar data do not
favour this possibility. It is still possible that the neutrino
mass matrix at high scale corresponds to  bi-maximal mixing and
radiative corrections lead \cite{cpv,lindner,miyura} to departure
from this.

We investigate this possibility in the present context assuming
maximal solar mixing and  a degenerate neutrino mass pair with
masses $m$ and $-m$ in a CP conserving theory at a high scale. The
maximal solar mixing automatically implies vanishing $U_{e3}$ and
thus our formalism applies to this case without further
assumptions. For zero $U_{e3}$ and equal and opposite masses, the
high scale solar mixing angle $\theta_0$ is given by

\be \tan^2\theta_0={m-\mee\over m+\mee}~, \ee
It is seen that the assumption of maximal mixing corresponds to
assuming that $\mee$ is zero at a high scale. The standard
radiative corrections result \cite{radrev}  in flavour dependent
renormalization of each elements of the neutrino mass matrix. As a
consequence $m_{ee}$ (which is the 11 element of the neutrino mass
matrix) remains zero in low energy theory and the solar angle does
not receive any corrections at $\ord(\delta_{\tau})$, see
eq.~(\ref{obs}). Thus solar mixing remains maximal to this order.
More importantly, vanishing of $m_{ee}$ implies through
eq.(\ref{pred}) that the predicted solar scale arise only at
$\ord(\delta_{\tau}^2)$ and hence is generically suppressed
 and can at most correspond to the vacuum solution with values of
$\delta_\tau\sim 10^{-5}$, ruling out the possibility of maximal
mixing and the LMA solution in the present context. CP
conservation assumed here is crucial. If CP violating Majorana
phases are included then the  maximal solar mixing at a high scale
does not necessarily imply vanishing $m_{ee}$. The maximal high
scale mixing becomes a viable possibility in this case \cite{cpv}.

\section{Textures and Models}
Eq. (\ref{cond}), leading to a degenerate pair and vanishing
$U_{e3}$ allow for a large number of models for  neutrino masses.
These conditions presuppose non-trivial relations among elements
of the neutrino mass matrix. It might therefore be thought that
such relations would represent a fine-tuned possibility and cannot
easily be obtained from a symmetry. This is not so. We present
here explicit textures which make these relations quite natural
and show how these textures can be obtained within a 
gauge-theoretical framework.

\subsection{Textures}\label{textures}

The first texture we discuss was postulated \cite{barb} long ago
 and is
also argued to emerge from a symmetry
\cite{bh}. This corresponds to the following neutrino mass matrix
in the flavour basis:
\bea \label{barb} M_{\nu_0}&=&m R_{23}(\theta_2) \bmat{ccc}
0&1&0\\
1&0&0\\
0&0&z\\ \emat R_{23}^T(\theta_2)
~, \nonumber \\
&=&m  \bmat{ccc} 0&c_2&-s_2\\
c_2&s_2^2 z&s_2 c_2 z\\
-s_2&s_2 c_2 z&c_2^2 z \\
\emat \eea
$R_{23}(\theta_2)$ denotes here a rotation in the 23 plane
with an angle $\theta_2$; $s_2=\sin\theta_2, c_2=\cos \theta_2$
and $z=\frac{m'}{m}$. $z$ is assumed real. The above mass
matrix satisfies the general conditions in eq.~(\ref{cond}) and
therefore leads to vanishing CHOOZ angle and a degenerate pair.
The 11 element of this matrix is zero. Thus it provides a specific
example of the general models with maximal mixing discussed in the
previous section. As eq.~(\ref{pred}) shows, one has vanishing
solar scale to leading order in radiative corrections $\delta_\t$.
The $\solm$ is however generated at higher order in $\delta_\t$
and was shown \cite{barb} to be
\be \label{barbs} \solm\approx -{m^3\over m'-m} \delta_\tau^2
\sin^2 2 \theta_A \ee
Remembering that $\delta_\tau^2 \approx 10^{-10}$ in the case of
SM, we see that the above texture leads to the vacuum solution of
the solar neutrino problem in general. Note however that there is
an enhancement by a factor $\frac{m^3}{m'-m}\approx {2 m^4 \over
\atm}\approx 10^3 ({m\over \eV})$ in eq.(\ref{barbs}) which partly
compensates for the additional power of $\delta_{\tau}$ but still
one does not get the LMA solution. Depending on the sign of
$m'-m$, now one can  get $\tan^2\sola<1$ in case of the MSSM
also. This can lead to additional enhancement for large
$\tan\beta$. Thus one may get the required solar scale for some
ranges in the parameter space for this model. However, the
predicted low scale mixing is nearly maximal which is not preferred in
case
of the LMA solution.

One possible way of generating the texture in eq.~(\ref{barb}) is
to have  a charged-lepton mass matrix which provides the
atmospheric mixing angle $s_2$ corresponding to the rotation
$R_{23}$ in eq.~(\ref{barb}). The neutrino mass matrix then will
be block diagonal with its upper block coinciding with the generic
pseudo-Dirac structure of eq.~(\ref{dirac}). This structure is
responsible for maximal solar mixing and vanishing $\mee$ which
results in a suppressed $\solm$. These problems can be
circumvented if the upper block of the neutrino mass matrix
corresponds to the generic pseudo-Dirac structure of
eq.~(\ref{pd2}) instead of eq.~(\ref{dirac}). This leads to the
the following texture:
\bea \label{pd3} M_{\nu 0}&=& R_{23}(\theta_2) \bmat{ccc}
a&b&0\\
b&-a&0\\
0&0&m'\\ \emat R_{23}^T(\theta_2)
~, \nonumber \\
&=& \bmat{ccc} a&b c_2&-b s_2\\
b c_2&-a c_2^2+m's_2^2 &s_2 c_2 (a+m')\\
-b s_2&s_2 c_2 (a+m')&-a s_2^2+m' c_2^2  \\
\emat \eea
This is a simple $3\times 3$ generalization of eq.~(\ref{pd2}). It
satisfies  the basic conditions in eq.(\ref{cond}) since  the above
$M_{\nu 0}$ is
constructed to yield a degenerate pair and vanishing $U_{e3}$. The
above texture represents a fairly general solution to  the
conditions of eq.~(\ref{cond})  since $M_{\nu 0}$ in eq.~(\ref{pd3})
contains maximum allowed ( namely four) independent parameters
after imposition of eq.~(\ref{cond}). The $\solm$ as given by
eq.~(\ref{pred}) and a non-zero $U_{e3}$ at the low scale are the
predictions of
this texture. Phenomenological consequences of these are already
elaborated in \cite{pdp2}. In particular one  gets the LMA
solution for $\mee=a\sim 0.2-0.4  \eV$.

\subsection{Gauge-theoretical realization}

We now discuss how the above textures can be realized in gauge
theories. We restrict ourselves  to a
non-supersymmetric theory, with the gauge group $SU(2)_L\times
U(1)_Y$ so far as the local symmetry is concerned. However to
generate the required pattern of neutral and charged-lepton mass
matrices, we impose global invariance under a ``horizontal" or
generation symmetry group\footnote{A local horizontal $SU(2)_H$ 
symmetry has
also been used recently in \cite{mk} to explain the bi-maximal
pattern.} $SU(2)_H \times U(1)_H$. Since we will employ the seesaw
mechanism for generation of small masses for the neutrinos, we
introduce three neutral right-handed $SU(2)_L$-singlet neutrinos,
$\nu_{eR}$, $\nu_{\mu R}$, $\nu_{\tau R}$. In order to generate a
Majorana mass matrix among these right-handed neutrinos, we need
to introduce $SU(2)_L$-singlet scalars $\eta$ and $T$. The number
of $SU(2)_L$-doublet scalars is increased to 4, viz., $\varphi$,
$\chi_1$, $\chi_2$, and $\phi$. These scalars all belong to the
doublet representation of $SU(2)_L$, but differ in their $SU(2)_H
\times U(1)_H$ transformation properties. The transformation
properties of the leptons and the scalars are given in Table
\ref{tra}.

\begin{table}[htb]
\begin{center}
\begin{tabular}{|c|cccc|}
\hline
        & $SU(2)_L$ & $U(1)_Y$ & $SU(2)_H$ & $U(1)_H$ \\
\hline
($l_e, l_\mu$) &2&   $-\frac{1}{2}$     & 2 & ~0 \\
$l_\tau$ & 2 & $-\frac{1}{2}$ & 1 & ~1 \\
($\nu_{eR}, \nu_{\mu R}$)&1&~0 & 2 & ~0 \\
$\nu_{\tau R}$&1& ~0           & 1 &$-1$\\
($e_R,\mu_R$)&1& $-1$         & 2 & ~0 \\
$\tau_R$ & 1 & $-1$           & 1 & $-1$ \\
\hline
$\varphi$& 2 & $-\frac{1}{2}$ & 1 & ~0 \\
$\chi_1$ & 2 & $-\frac{1}{2}$ & 1 &$-2$ \\
$\chi_2$ & 2 & $-\frac{1}{2}$ & 1 &$+2$\\
$\phi$   & 2 & $-\frac{1}{2}$ & 2 &$-1$ \\
$T$      & 1 & ~0              & 3 & ~0   \\
$\eta$   & 1 & ~0 & 1 & ~2 \\
\hline
\end{tabular}
\end{center}
\caption{Transformation properties of lepton and scalar fields under the gauge
group $SU(2)_L\times U(1)_Y$ and the horizontal symmetry group $SU(2)_H \times
U(1)_H$}
\label{tra}
\end{table}

With the transformation properties given in Table \ref{tra}, the most
general Yukawa
couplings of the leptons with the various scalar fields consistent with the
symmetries can be written as
\bea\label{yuk}
- {\cal L}_Y &=& h_1\, ( \o l_e e_R  + \o l_\mu \mu_R) \tilde \varphi
        + h_2\, \o l_\tau (\tilde \phi_2 \mu_R - \tilde \phi_1 e_R )
            \nonumber \\ &&
        + h_3\, (\o l_e \tilde \phi_2 + \o l_\mu \tilde \phi_1)\tau_R
        + h_4\, (\o l_\tau \tau_R \tilde \chi_1)
            \nonumber \\ &&
        + h'_1\, (\o l_e \nu_{eR} + \o l_\mu \nu_{\mu R})\varphi
        + h'_4\, \o l_\tau \nu_{\tau R} \chi_2
            \nonumber \\ &&
        +ik\, (\begin{array}{lr} \nu^T_{eR} & \nu^T_{\mu R} \end{array})
            \tau_2 \vec{T}\cdot \vec{\tau}
        \left( \begin{array}{c} \nu_{eR} \\ \nu_{\mu R} \end{array}
            \right) \nonumber \\ &&
        + k'\, \nu^T_{\tau R} \nu_{\tau R} \eta +{\rm  H.c.}
\eea
In the above equation, the notation is that $\phi$, which is a doublet
under $SU(2)_L$ as well as $SU(2)_H$, is written as
\be
\phi \equiv ( \begin{array}{lr} \phi_1 & \phi_2 \end{array} ),
\ee
where the $SU(2)_L$ doublets $\phi_1$ and $\phi_2$ have the components
\be
\phi_1 \equiv \left( \begin{array}{cc} \phi_1^0 \\ \phi_1^- \end{array} \right)
\ee
and
\be
\phi_2 \equiv \left( \begin{array}{cc} \phi_2^0 \\ \phi_2^- \end{array}
\right).
\ee
The field $\tilde \phi$ conjugate to $\phi$ is given by
\be
\tilde \phi = \tau_2 \phi^* \tau_2
        = \left( \begin{array}{rr} \phi_2^+ & - \phi_1^+ \\
                      -\phi_2^{0*} & \phi_1^{0*}
                \end{array} \right)
        \equiv ( \begin{array}{lr} \tilde \phi_2 & \tilde \phi_1
                \end{array} ).
\ee

Since only electrically neutral components of the Higgs scalars are permitted
to have nonzero vacuum expectation values (vev's), we first separate the Yukawa
terms involving the neutral scalars:
\bea\label{neu}
- {\cal L} & = & h_1\, (\o e_L e_R  + \o \mu_L \mu_R ) \varphi^{0*} +
        h_2 \,\o \tau_L (-\phi_2^{0*} \mu_R - \phi_1^{0*} e_R )
        \nonumber \\ &&
        + h_3\, ( - \o e_L \phi_2^{0*} + \o \mu _L \phi_1^{0*}) \tau_R
        + h_4\, \o \tau_L \tau_R \chi_1^{0*}
        \nonumber \\ &&
        + h'_1\, ( \o \nu_{eL} \nu_{eR} + \o \nu_{\mu L} \nu_{\mu R})
        \varphi^0 + h'_4\, \o \nu_{\tau L} \nu_{\tau R} \chi_2^0
        \nonumber \\ &&
        + k\, ( \ba{cc} \nu^T_{eR} & \nu^T_{\mu R} \ea )
        \left( \ba{cc}  T^1 -i T^2 & - T^3 \\
            -T^3 & -( T^1 +i T^2) \ea \right)
            \left( \ba{cc} \nu_{eR} \\ \nu_{\mu R} \ea \right)
            \nonumber \\ &&
        + k'\, \nu^T_{\tau R}\nu_{\tau R} \eta + {\rm H.c.}
\eea

When the neutral components of the scalar fields are replaced by their vacuum
expectation values which arise due to spontaneous symmetry breaking, we get
the mass terms of the charged and neutral leptons from (\ref{neu}). The
vacuum expectation values for the neutral fields are defined as
$$
\langle \varphi^0 \rangle \equiv v; \;\;
    \langle \chi_1^0 \rangle \equiv \kappa_1; \;\;\langle \chi_2^0 \rangle
    \equiv \kappa_2;
$$
\be\label{vevs}
\langle T^1 \rangle \equiv v_R; \;\;\langle T^3 \rangle \equiv v'_R ;
\ee
$$
\langle \phi^0_1 \rangle  \equiv v_1; \;\;\langle \eta \rangle \equiv
v_\eta.
$$
Note that we have set the vacuum expectation values for two of the
neutral fields namely, $\phi^0_2$ and $T_2$ to be zero. Using the
invariance of the original Lagrangian under $SU(2)_L$ and
$SU(2)_H$ one can always make $\langle \phi_2^0\rangle$ zero
without loss of generality\footnote{ We are assuming that the
Higgs potential parameters are such that there is no charge
breaking minima in this basis with zero $\langle
\phi_2^0\rangle$.}. The choice $\langle T_2 \rangle=0$ follows if
we demand CP invariance. Terms involving triplet in Eq. (\ref{neu})
are invariant under CP for a real $\kappa$ if $T^1+iT^2\rightarrow
T^1-iT^2$, i.e. $(T^1,T^2)\rightarrow (T^1,-T^2)$ under CP.
Requiring that vacuum also respects CP, one gets zero vacuum
expectation value for $T^2$. This can always be done by proper
choice of parameters in the potential. We have allowed non-zero
and real vacuum expectation values for all other neutral fields in
the model.

With the choice of eq.~(\ref{vevs}) for the vev's, eq.~(\ref{neu}) gives rise
to the following mass matrices. The charged-lepton mass matrix is
\be\label{mch}
M_l = \left( \ba{ccc}
            m_{11}   & 0      & 0      \\
            0     & m_{22}    & m_{23}   \\
            m_{31}     & 0 & m_{33}
    \ea \right),
\ee
where
\be
m_{11} = m_{22} = h_1 v; \;\; m_{23} = h_3 v_1; \;\; m_{31} = -h_2
v_1; \;\;
        m_{33} = h_4 \kappa_1 .
\ee
Note that the $2-3$ block of the charged lepton mass
matrix
has the lopsided \cite{lop} texture which has been advocated
as an explanation of the large atmospheric mixing \cite{barr}.

The neutral-lepton mass matrix may be written as
\be\label{mneu}
M_N = \left( \ba{cc} 0 & m_D \\
             m_D^T & M_R \ea \right),
\ee
where
\be\label{mD}
m_D = \left( \ba{ccc}  \m & 0 & 0 \\
               0 & \m & 0 \\
               0 & 0 & \m' \ea \right),
\ee
and
\be\label{mR}
M_R = \left( \ba{crc} M   & M_1 & 0 \\
              M_1 & -M  & 0 \\
              0   &  0  & M'  \ea \right),
\ee
with
\be
\m = h'_1 v; \;\; \m' = h'_4 \kappa_2
\ee
and
\be
M = 2 k v_R; \;\; M_1 = -2 k v'_R; \;\; M' = 2 k' v_\eta.
\ee

The neutral lepton mass matrix $M_N$ can be block diagonalized as usual in the
seesaw approximation,
\be
\{\vert \m \vert , \vert \m' \vert \} \ll \{ \vert M \vert , \vert M_1
\vert ,
                    \vert M' \vert \}.
\ee
The resulting mass matrix for the light neutrinos is, in the seesaw limit,
\be\label{mnu}
m_{\nu}\! =\! - m_D^T M_R^{-1} m_D\! =\! \frac{1}{M' (M^2 + M^2_1)}
            \!  \left(\!\!\!\! \ba{rrc} -\m^2 M  & -\m^2 M_1 & 0 \\
                        -\m^2 M_1&  \m^2 M   & 0 \\
                           0  &    0  &-\m'^2(M^2+M^2_1)
                    \ea \right).
\ee
This has the form
\be\label{mnu2}
m_{\nu} = \left( \ba{rrc} a & b & 0 \\
              b &-a & 0 \\
              0 & 0 & m'   \ea \right),
\ee 
identical to the one discussed in Sec.~\ref{textures}.
As discussed earlier, it
gives rise to a pseudo-Dirac neutrino, and implies large mixing
between the first two generations, for  values of $a$ and $b$
comparable in magnitude. 

The charged-lepton  mass matrix leads to an additional $23$ rotation,
which generates the atmospheric mixing angle. 
Eq.(\ref{mch}) can be used to obtain
\be \label{mlmld} M_l M_l^\dagger=\bmat{ccc} 
m_{11}^2&0&m_{11}m_{31}\\
0& m_{11}^2+m_{23}^2& m_{23} m_{33} \\
m_{11} m_{31}&m_{23}m_{33}& m_{13}^2+m_{33}^2\\ \emat \ee
The above matrix leads to the required charged lepton masses
with $m_{11}\sim \ord(m_e)$, $m_{23}\sim m_{33}\sim\ord(m_\tau)$ and
$m_{31}\sim \ord(m_\mu)$.  Eq.(\ref{mlmld}) can be diagonalized
in the limit $m_e\rightarrow 0$ by a rotation $R_{23}(\theta_2)$ 
in the $2-3$ plane with the mixing angle $\theta_2$ given by
\be
\tan 2 \theta_{2} =  \frac{ 2 (m_{23} m_{33})}
            { m^2_{33} + m^2_{13} - m^2_{23}}.
\ee
This angle is large for the values of parameters quoted above, which
correctly reproduce the charged lepton masses. In this way, eqs.(\ref{mch}) 
and (\ref{mnu2}) reproduce the texture of eq.(\ref{pd3})
in the limiting case of zero electron mass. The effect of introducing
a non-zero $m_{11}\sim m_e$ on the mixing is negligible. As seen from
eq.(\ref{mlmld}), a non-zero $m_{11}$  leads to a  $U_{e3}$ at the
high scale
$$ U_{e3} \approx {m_e m_{\mu}\over m_{\tau}^2}\approx 10^{-5} $$
This is negligible and does not effect the phenomenological consequences
discussed in this paper.

\section{Conclusions}

We have examined the consequences of a general $3\times 3$ flavour-basis
neutrino mass matrix which satisfies the conditions necessary for two
eigenvalues to be equal in magnitude and opposite in sign, wherein
radiative corrections lead to a pseudo-Dirac structure, and in addition, which
leads to a vanishing $U_{e3}$ matrix element at tree level. CP
conservation was assumed. As shown earlier \cite{pdp2}, these assumptions
lead to a strong correlation between the neutrino mass $m_{ee}$
determined from \dbd experiments and the solar scale.
Specific testable phenomenological predictions of the scenario considered
here are very
small
($\leq .01$) $U_{e3}$, and $m_{ee}$ as well as the electron neutrino mass
close to the present experimental limit if the LMA solution is verified.

General important theoretical results shown to follow from this scenario
were that
(i) radiative corrections in MSSM
are inconsistent with the requirement of $\tan^2\sola\leq 1$ demanded by the
LMA and LOW solutions of the solar neutrino problem, (ii) the inverted mass
hierarchy, realized as for example in models with $L_e - L_\mu - L_\tau$
symmetry, cannot lead to the LMA solution, and (iii) maximal solar mixing at
tree level, which is simple from a conceptual and theoretical point of view,
does not decrease appreciably after radiative corrections, and is therefore
ruled out in the context of the LMA solution. In the last case, even the
pseudo-Dirac splitting is too small, except possibly for the vacuum solution
for solar neutrinos.

The stringent relations among the elements of the neutrino mass matrix for the
scenario do not necessarily require fine-tuning of parameters.
They are shown in a couple of examples to follow from some simple textures. A
gauge-theoretic realization of such a texture was also explicitly
worked out.

\end{document}